\documentstyle[epsf,12pt]{article}
\newcommand {\beq}{\begin{equation}}
\newcommand {\eeq}{\end{equation}}
\newcommand {\beqa}{\begin{eqnarray}}
\newcommand {\eeqa}{\end{eqnarray}}
\newcommand {\n}{\nonumber \\}

\renewcommand{\theequation}{\thesection.\arabic{equation}}
\begin{document}
\setlength{\oddsidemargin}{0cm}
\setlength{\baselineskip}{7mm}

\begin{titlepage}
 \renewcommand{\thefootnote}{\fnsymbol{footnote}}
$\mbox{ }$
\begin{flushright}
\begin{tabular}{l}
KEK-TH-874\\
Mar. 2003
\end{tabular}
\end{flushright}

~~\\
~~\\
~~\\

\vspace*{0cm}
    \begin{Large}
       \vspace{2cm}
       \begin{center}
         {Quantum Corrections on Fuzzy Sphere}      \\
       \end{center}
    \end{Large}

  \vspace{1cm}

\begin{center}
           Takaaki I{\sc mai}$^{2)}$\footnote
           {
e-mail address : imaitakaaki@yahoo.co.jp},
           Yoshihisa K{\sc itazawa}$^{1),2)}$\footnote
           {
e-mail address : kitazawa@post.kek.jp}\\
           Yastoshi T{\sc akayama}$^{2)}$\footnote
           {
e-mail address : takaya@post.kek.jp}{\sc and}
           Dan T{\sc omino}$^{1)}$\footnote
           {
e-mail address : dan@post.kek.jp}

        $^{1)}$ {\it High Energy Accelerator Research Organization (KEK),}\\
               {\it Tsukuba, Ibaraki 305-0801, Japan} \\
        $^{2)}$ {\it Department of Particle and Nuclear Physics,}\\
                {\it The Graduate University for Advanced Studies,}\\

{\it Tsukuba, Ibaraki 305-0801, Japan}\\
\end{center}

\vfill

\begin{abstract}
\noindent
We investigate quantum corrections in non-commutative gauge theory on fuzzy
sphere. We study translation invariant models which
classically favor a single fuzzy sphere with $U(1)$ gauge group.
We evaluate the effective actions up to the two loop level.
We find non-vanishing quantum corrections at each order
even in supersymmetric models.
In particular the two loop contribution favors
$U(n)$ gauge group over $U(1)$ contrary to the tree action
in a deformed IIB matrix model with a Myers term.
We further observe close correspondences to 2 dimensional quantum gravity.
\end{abstract}
\vfill
\end{titlepage}
\vfil\eject

\section{Introduction}
\setcounter{equation}{0}
Matrix models are a promising approach to understand
non-perturbative aspects of superstring/M-theory\cite{BFSS}\cite{IKKT}.
In this approach, space-time and matter fields may emerge
from matrix degrees of freedom.
A well-known prototype of such a possibility is the following
non-commutative D-brane solutions in the
large $N$ limit
\beqa
&&A^{cl}_1=\hat{p},~~A^{cl}_2=\hat{q},\n
&&[\hat{p},\hat{q}]=-i .
\eeqa
Non-commutative (NC) gauge theory is obtained from matrix models
around such solutions\cite{CDS}\cite{AIIKKT}\cite{Li}.
In string theory, NC gauge theory on flat space is realized with
constant $B_{\mu\nu}$ field\cite{SW}.
NC gauge theory exhibits UV-IR mixing which is
a characteristic feature of string theory\cite{MRS}.
The advantage of matrix model construction of NC gauge theory
is that it maintains the manifest gauge invariance under
$U(N)$ transformations
\beq
A_{\mu} \rightarrow UA_{\mu}U^{\dagger} .
\eeq
The gauge invariant observables of NC gauge theory,
the Wilson lines were constructed through matrix
models\cite{IIKK}\cite{Gross}.
In bosonic string theory, they are the emission vertex operators for
the entire closed string modes\cite{DK}\cite{MN}.

To further elucidate the dynamics of the matrix models
for superstring/M-theory,
it may be useful to investigate matrix models on homogeneous
spaces.
Experimentally it is now very likely that we live
in a homogeneous space, namely de-Sitter space.
A homogeneous space is realized as $G/H$
where $G$ is a Lie group and $H$ is a closed subgroup of $G$.
NC gauge theories on homogeneous spaces are realized by matrix models.
In string theory, they may appear with non-constant
$B_{\mu\nu}$ field\cite{Myers}\cite{Alekseev}.
It is an interesting problem on its own to
study NC gauge theories on curved manifolds.

We have formulated a general procedure to construct
fuzzy homogeneous spaces $G/H$ in \cite{Mathom}.
We first consider a representation of $G$ which contains
a (highest weight) state which is invariant under $H$ modulo
a local gauge group.
We also require that fuzzy $R^n$ where $n$ is the dimension of $G/H$
is realized in the local patch around such a state.
We are thus restricted to symplectic manifolds. It is because
the $\star$ product on such a manifold can be
reduced to that of a flat manifold (Moyal product) locally by choosing
the Darboux coordinates.
Since the K\"{a}hler form serves as the symplectic form,
K\"{a}hler manifolds such as $CP^n$ satisfy this requirement\cite{Masuda}.

We embed the Lie generators of $G$ into
$N$ dimensional Hermitian matrices where $N$ is the
dimension of the representation.
The gauge fields on fuzzy $G/H$ are constructed as bi-local fields\cite{IKK}.
The bi-local fields are the tensor products
of the relevant representation and the complex conjugate of it.
Since they are reducible, we can decompose them into the irreducible
representations.
They are guaranteed to form the complete basis of $N\times N$
Hermitian matrices by construction.

In this paper we investigate quantum aspects of
matrix models on fuzzy $S^2$ up to the two loop level.
Although $S^2$ has been chosen for simplicity, it may be
possible to draw implications for generic homogeneous spaces
from it.
We can address such physically interesting questions
as the SUSY breaking effects
and correspondences to quantum gravity in a concrete setting.

Let us briefly summarize the basic facts of fuzzy $S^2$.
$S^2$ is a homogeneous space since $S^2=SU(2)/U(1)$.
In $SU(2)$, we have the Hermitian operators
$\hat{j}_x,\hat{j}_y,\hat{j}_z$ which satisfy the commutation relations
of the angular momentum:
\beq
[\hat{j}_x,\hat{j}_y]=i\hat{j}_z .
\eeq
Contrary to $R^2$ case, such a commutation relation
can be realized with finite size matrices
since $S^2$ is compact.
The raising and lowering operators $\hat{j}^+,\hat{j}^-$ can be formed from
$\hat{j}_x,\hat{j}_y$ which satisfy
\beqa
&&\hat{j}^{\pm}={1\over \sqrt{2}}(\hat{j}_x\pm i\hat{j}_y),\n
&&[\hat{j}^+,\hat{j}^-]=\hat{j}_z ,
~~[\hat{j}_z, \hat{j}^{\pm}]=\pm \hat{j}^{\pm} .
\label{angcom}
\eeqa
Let us adopt an $N=2l+1$ dimensional
representation of spin $l$.
We further consider the semiclassical limit where $l$ is assumed to be
large.
For the localized states around the north pole,
we can approximate $\hat{j}_z \sim l$.
By rescaling the operators, we obtain the following
commutation relations from (\ref{angcom}),
\beqa
&&\tilde{a}={1\over \sqrt{l}}\hat{j}^+,
~\tilde{a}^{\dagger}={1\over \sqrt{l}}\hat{j}^-,
~\tilde{1}={1\over l}\hat{j}_z\n
&&[\tilde{a},\tilde{a}^{\dagger}] = \tilde{1},
~~[\tilde{1},\tilde{a}]={1\over l}\tilde{a},
~~[\tilde{1},\tilde{a}^{\dagger}]=-{1\over l}\tilde{a}^{\dagger} .
\label{s2com}
\eeqa
Since we obtain the identical algebra with fuzzy $R^2$,
flat fuzzy plane is locally realized.
Note that the radius of $S^2$ is $\sqrt{l}$
after fixing the non-commutativity scale to be 1.

The gauge fields are constructed as the bi-local fields.
They are the tensor product of the spin $l$ representations
\beq
\b{l}~\otimes \b{l}
=\sum_{j=0}^{2l}~ \b{j} .
\eeq
From the above decomposition, we can see that
a group of representations with spins up to $2l$
form the complete basis of $N\times N$ Hermitian matrices.

The organization of this paper is as follows.
We initiate our investigation of quantum effects in NC gauge theory
on fuzzy sphere through matrix models
in section 2.
We summarize our results up to the two loop level
and discuss their physical implications in section 3.
We conclude in section 4 with discussions.
We delegate the detailed evaluation of the two loop effective action
to Appendix A.

\section{Quantum effects in matrix models}
\setcounter{equation}{0}

NC gauge theories on compact homogeneous spaces
can be constructed through matrix models.
A minimal model which realizes NC gauge theories
on fuzzy sphere is constructed in\cite{IKTW}.
It is a reduced super Yang-Mills theory in 3d
with a Myers term:
\beq
S=Tr\left(-{1\over 4}[A_{\alpha},A_{\beta}][A_{\alpha},A_{\beta}]+
{i\over 3}f\epsilon^{\alpha\beta\gamma}[A_{\alpha},A_{\beta}]A_{\gamma}
+{1\over 2}\bar{\psi}\sigma^{\alpha}[A_{\alpha},\psi]\right) ,
\label{isoact}
\eeq
where $A_{\alpha}$ and $\psi$ are $N\times N$ Hermitian matrices.
$\psi$ is a two component Majorana spinor field
and $\sigma^{\alpha}$ are Pauli matrices.
The action is invariant under the following supersymmetry:
\beqa
&&\delta A_{\alpha} = i\bar{\epsilon}\sigma_{\alpha}\psi ,\n
&&\delta \psi = {i\over 2}\left([A_{\alpha},A_{\beta}]-
if\epsilon_{\alpha\beta\gamma}A_{\gamma}\right)
\sigma^{\alpha\beta}\epsilon .
\label{susytr}
\eeqa

We can also deform IIB matrix model in an analogous way
\cite{Mathom}
\beq
S_{IIB}\rightarrow S_{IIB}+
{i\over 3} f_{\mu\nu\rho}Tr[A_{\mu},A_{\nu}]A_{\rho} ,
\label{IIBdef}
\eeq
where $f_{\mu\nu\rho}$ is the structure constant of a compact Lie group $G$.
The perturbation term may represent non-constant $B_{\mu\nu}$ field
\cite{Kitazawa}.
Since there are 10 Hermitian matrices $A_{\mu}$ in IIB matrix model,
the number of the Lie generators of $G$ cannot exceed 10 in this
construction.
Although this model no longer preserves SUSY, it
possesses many similarities to
(\ref{isoact}) when $G=SU(2)$.
\footnote
{Although it is possible to construct a deformed IIB matrix
model with SUSY when $G=SU(2)$\cite{Bonel},
fuzzy $S^2$ solutions break SUSY at the classical level in that
model.}

Since these models possess the translation invariance
\beq
A_{\mu}\rightarrow A_{\mu}+c_{\mu}
\eeq
and also
\beq
\psi\rightarrow \psi+\epsilon,
\eeq
we remove these zero-modes by restricting $A_{\mu}$ and $\psi$
to be traceless.

The equation of motion is
\beq
[A_{\mu},[A_{\mu},A_{\nu}]]+i f_{\mu\rho\nu}[A_{\mu},A_{\rho}]=0 .
\eeq
The nontrivial classical solution is
\beq
A_{\alpha}^{cl}= t^{\alpha},
~~other ~A_{\mu}^{cl} =0 ,
\label{clsol}
\eeq
where $t^{\alpha}$'s satisfy the Lie algebra of $G$.

We consider the following classical solution for $G=SU(2)$ case.
\beq
A_{\alpha}^{cl}=f j_{\alpha}\otimes 1_{n\times n} ,
~~A^{cl}_{i}=0 ,
\eeq
where $j_{\alpha}$ are angular momentum operators
in the spin $l$ representation and $i$ denotes
the orthogonal directions to the 3 dimensional space in which $S^2$ resides.
The fixed parameters of the matrix models are $N$ (the dimension of the
matrices) and $f$ (the coefficient of the Myers term).
Since different $U(n)$ gauge groups could be realized
as long as $N$ is divisible by $n$, gauge groups
are dynamically determined in the matrix models.
The classical action associated with this solution is
\beq
-{f^4\over 6}nl(l+1)(2l+1) .
\label{clsact}
\eeq
For a large but fixed $N=2l_{max}+1$, the irreducible representation
of spin $l_{max}$ is selected by
minimizing the classical action\cite{Myers}.

In the large $N$ limit, the fuzzy $R^2$ may be realized locally
as we recalled it in the preceding section.
The local momenta are introduced as $j_{\alpha}=\sqrt{l}\hat{p}_{\alpha}$.
We expand the action around the classical solution as
$A_{\alpha}=f\sqrt{l}(\hat{p}_{\alpha}+\hat{a}_{\alpha}),
A_{i}=f\sqrt{l}\phi_i$.
In this paragraph, $\alpha$ denotes the tangential directions to $R^2$
while $i$ denotes the transverse directions to it.
Here we have fixed the non-commutativity scale
to be 1.
After using the Moyal-Weyl correspondence,
\beqa
&&\hat{a}\rightarrow a(x),\n
&&\hat{a}\hat{b}\rightarrow a(x)* b(x),\n
&&Tr\rightarrow ({1\over 2\pi})\int d^2x tr,
\eeqa
we obtain the following NC gauge theory from (\ref{IIBdef})
\beqa
&&-f^4 l^2({1\over 2\pi})\int d^2x
tr\left({1\over 4}[D_{\alpha},D_{\beta}]^2
+{1\over 2}[D_{\alpha},\phi_i]^2
+{1\over 4}[\phi_i,\phi_j]^2\right.\n
&&\left.+{1\over 2}\bar{\psi}\Gamma_{\alpha}[D_{\alpha},\psi]
+{1\over 2}\bar{\psi}\Gamma_{i}[\phi_i,\psi] \right)_{*} .
\label{nc2act}
\eeqa
$tr$ denotes the trace operation over $U(n)$ gauge group.
The Chern-Simons terms are suppressed by $1/\sqrt{l}$.
In this way, we identify the coupling constant of NC gauge theory
at the non-commutativity scale as
\beq
g^2_{NC_2}=2\pi ({1\over lf^2})^2.
\label{2dcoup}
\eeq
The classical action (\ref{clsact}) is $O(N)$ for a finite gauge coupling
\beq
-{2\pi nl\over 3g_{NC}^2}\sim -{2\pi N\over 6g_{NC}^2} .
\eeq
In order to obtain a finite gauge coupling $g_{NC}$, we need to choose
$f^2 \sim 1/l$.
Since we find $N\sim O(l)$ in 2 dimensions, we need to let $f$
vanish in the large $N$ limit as $f \sim O(1/\sqrt{N})$.
From (\ref{nc2act}), we can see that SUSY is locally recovered
in this limit.

In order to obtain NC gauge theory,
we expand the matrices around the classical solution:
\beq
A_{\alpha}=f (j_{\alpha}+a_{\alpha}),
~~A_{i}=f \phi_i~~(i=1\sim 7),
~~\psi\rightarrow f^{3\over 2}\psi .
\eeq
Here we have separated bosonic fluctuations into vector (triplet)
$a_{\alpha}$ and scalar (singlet) $\phi_i$ fields.
After this procedure, the action (\ref{IIBdef}) becomes
\beqa
&&f^4 Tr
\left[ -{1\over 4}
(L_{\alpha} a_{\beta}-L_{\beta}
a_{\alpha}+[a_{\alpha},a_{\beta}])^2
-{1\over 2}
(L_{\alpha}\phi_i+[a_{\alpha},\phi_i])^2
-{1\over 4}[\phi_i,\phi_j]^2 \right. \n
&&+\left.{i\over 2}\epsilon_{\alpha\beta\gamma}
(L_{\alpha} a_{\beta}-L_{\beta} a_{\alpha})a_{\gamma}
+{i\over 3}\epsilon_{\alpha\beta\gamma}
[a_{\alpha}, a_{\beta}]a_{\gamma}  \right.\n
&&\left.+
{1\over 2}
\bar{\psi}\Gamma^{\alpha}(L_{\alpha}\psi+[a_{\alpha},\psi])
+{1\over 2}\bar{\psi}\Gamma^{i}[\phi_i,\psi ]
\right] .
\label{expact}
\eeqa
where $L_{\alpha}X\equiv [j_{\alpha},X]$.

After the gauge fixing,
the bosonic quadratic action is simply given by
\beq
{f^4\over 2}
Tr\left({a}_{\alpha}L^2{a}_{\alpha}
+\phi_iL^2\phi_i
\right) .
\eeq
$L^2$ is the Laplacian on fuzzy $S^2$
which acts on matrix spherical harmonics $Y_{jm}$ as
\beq
L^2Y_{jm}=j(j+1)Y_{jm} .
\eeq
In our matrix model, $j$ is constrained as $1\leq j \leq 2l$.
We thus expand bosonic quantum fluctuations in terms of $2l+1$ dimensional
matrix spherical harmonics $Y_{jm}$:
\beq
a=\sum_{jm}Y_{jm}a_{jm},
~~\phi=\sum_{jm}Y_{jm}\phi_{jm} ,
\eeq
where the expansion coefficients $a_{jm},\phi_{jm}$ are
$n$ dimensional Hermitian matrices in turn.

The modes with spin $j=0$ are zero-modes since the inverse
propagators vanish for them. There are $n^2-1$ of them since we have
removed the trace part already using the translation invariance.
They are the moduli of $n$ coincident fuzzy spheres.
Our strategy is to  integrate massive modes first to obtain the
Wilsonian effective action.
Apart from the constant term which measures the free
energy of the local vacuum,  it is also the functional of zero-modes.
We argue that it must be a non-polynomial matrix model of $n\times n$
Hermitian matrices which govern the dynamics of the moduli fields.
In this paper we compute the constant part of the effective
action up to the two loop level since it is an
important step to understand the dynamics of the model.
We are content to briefly discuss one loop effective action for
zero modes.

The quadratic action for $\psi$ is
\beq
{1\over 2}\bar{\psi}\Gamma^{\alpha}L_{\alpha}\psi .
\eeq
In the 3d model of (\ref{isoact}), we can adopt
\beq
\Gamma^0=\sigma^2,
~~\Gamma^1=\sigma^3,
~~\Gamma^2=\sigma^1 .
\eeq
For Majorana-Weyl spinors in ten dimension,
we can effectively factorize $\Gamma$ matrices as
\beq
\Gamma^{\mu}=\tilde{\gamma}^0\tilde{\gamma}^{\mu} ,
\eeq
where
\beqa
&&\tilde{\gamma}^0=\sigma^2\otimes 1_{8},
~~\tilde{\gamma}^9=-\sigma^1\otimes 1_{8},
~~\tilde{\gamma}^8=\sigma^3\otimes 1_{8},\n
&&\tilde{\gamma}^i=-1_{2}\otimes \gamma^i_{8}
~~(1\leq i\leq 7) ,
\eeqa
and $\gamma^i_{8}$  are real and antisymmetric.
With the choice of $\epsilon^{\alpha\beta\gamma}=\epsilon^{098}=1$,
the fermionic kinetic term of (\ref{IIBdef})
boils down to that of (\ref{isoact})
with the multiplicity of 8.

Let us expand $\psi$ in terms of $Y_{jm}$
\beq
\psi=\sum_{jm}\psi_{jm}Y_{jm} .
\eeq
Since $\psi$ carries spin ${1\over 2}$,
the total angular momentum of $\psi_{jm}$ is
either $j+{1\over2}$ or $j-{1\over 2}$.
For each state,
the eigenvalue of $\sigma^{\alpha}L_{\alpha}$ is
either $j$ with the multiplicity of $2j+2$ or
$-(j+1)$ with the multiplicity of $2j$ respectively.
It is because
\beq
\sigma^{\alpha}L_{\alpha}=
(L_{\alpha}+{\sigma_{\alpha}\over 2})^2-L^2-{3\over 4} .
\eeq

In our calculation of the partition function, we divide out the
following gauge volume of $SU(N)/Z_N$ by gauge fixing
\beq
2^{{N^2+N\over 2}-1}\pi^{N-1\over 2}
{1\over \sqrt{N}}{1\over\prod_{k=1}^{N-1} k!}  ,
\eeq
which appeared as the universal factor in \cite{KNS}.
We recall that the `exact' free energy of IIB matrix model
is as follows in this normalization\cite{MNS}
\beq
-log(\sum_{n|N}{1\over n^2}) .
\eeq

We have 8 real bosonic degrees of freedom
after subtracting the ghost contribution.
We have also 8 $SU(2)$ doublet real fermionic degrees of freedom as we have
just seen. On the other hand, the degrees of freedom is one each in
(\ref{isoact}).
Since each field is an $n$ dimensional Hermitian matrix,
the one loop determinant takes the following form.
\beqa
&&\left((1\cdot 2)^3(2\cdot 3)^5\cdots (j-1\cdot j)^{2j-1}
(j\cdot j+1)^{2j+1}\cdots \right)^{-4n^2}\n
&&\left(1\cdot 2^2 \cdot 2^6\cdot 3^4 \cdots
(j-1)^{2j}\cdot j^{2j-2} \cdot j^{2j+2}\cdot (j+1)^{2j}
\cdots \right)^{4n^2} .
\eeqa

Let us focus on the multiplicity of the eigenvalue of $j$
in the above expression.
In the fermionic sector, the multiplicity is
$(2j+2)+(2j-2)=4j$. The multiplicity in the bosonic sector
is also $(2j+1)+(2j-1)=4j$. Therefore the bosonic and fermionic contributions
cancel in general.
However we need to recall that we have the upper cut-off for $j$ as
it cannot exceed $2l$.
Concerning the largest possible factor $2l+1$ in the one loop
determinant,
the bosonic and fermionic multiplicity is $4l+1$  and $4l$
respectively.
Since they do not match,
we find the following constant term in the effective action
\beq
4n^2log(2l+1) .
\label{cnst}
\eeq

At one loop level, we can obtain
the effective action with a generic background $p_{\mu}$
in a closed form as
\beqa
&&{1\over 2}Trlog(P^2\delta_{\mu\nu}
-2iF_{\mu\nu}-2i\epsilon_{\mu\nu\rho}P^{\rho})-Trlog(P^2)\n
&&-{1\over 4}Trlog\left( (P^2+{i\over 2}F_{\mu\nu}\Gamma^{\mu\nu})
({1+\Gamma_{11}\over 2})\right),
\label{1lpact}
\eeqa
where
\beqa
&&[p_{\mu},X]=P_{\mu}X,\n
&&[f_{\mu\nu},X]=F_{\mu\nu}X,~~f_{\mu\nu}=i[p_{\mu}, p_{\nu}].
\eeqa
With our choice of $p_{\mu}=j_{\mu}$, we find
$F_{\mu\nu}=- \epsilon_{\mu\nu\rho}L^{\rho}$.
In order to evaluate the contributions from large
eigenvalues,
we may expand this expression into
the power series of $F_{\mu\nu}$.
The leading correction is found to be
\beq
2Tr{1\over L^2}
=2n^2\sum_{j=1}^{2l}(2j+1){1\over j(j+1)}
\sim 4n^2log(2l) .
\label{cosmo}
\eeq
We find that it is consistent with
the constant term (\ref{cnst}).

The one loop quantum correction in the model of $(\ref{isoact})$
is precisely $1/8$ of this result.
We thus find non-vanishing one loop quantum corrections
even in a supersymmetric matrix model.
Since SUSY transformation of $\psi$ (\ref{susytr}) vanishes in these
backgrounds, one might expect that fuzzy spheres do not
receive quantum corrections. We observe that this expectation
failed because the bosonic and fermionic
degrees of freedom do not match at the upper cut-off.
This miss-match can be further traced to the fact that bosonic
fields $a_{\mu}$ are in the integer and
the fermionic fields $\psi$ are in
the half integer representation of $SU(2)$.

This fact might sound contradictory to the presence of
SUSY transformation (\ref{susytr}) in this model.
However we can argue that it must be broken since
there cannot be unbroken supersymmetry
in de-Sitter space\cite{deSit}.
Since there is no positive conserved energy in de-Sitter space,
there is no supercharge $Q$ whose square equals the Hamiltonian.
Since the Euclidean continuation of an $n$-dimensional de-Sitter space
$dS_n$ is an $n$ sphere $S^n$, our conclusion follows.
In other words, it is a thermal SUSY breaking effect
due to a de-Sitter temperature $O(1/l)$.
We further note that KMS conditions are satisfied since `spin-statistics'
relations are respected as we just mentioned.

Let us reflect on the one loop free energy from
quantum gravity point of view\cite{HLW}\cite{AK}.
The matter contribution to the geometric entropy
in 2d supergravity can be obtained through conformal anomaly
on $S^2$:
\beq
-{\delta S\over \delta \varphi}
={c\over 4} ,
\eeq
where $c$ denotes the matter central charge.
Semiclassically
we can integrate this equation as
\beq
-S=-{c\over 4}log(A) ,
\eeq
where $A$ is the ratio of short and long distance cut-offs.

On the other hand we may rewrite (\ref{cosmo}) as
\beq
-4n^2log(2l) +8n^2log(2l)
\eeq
The second term in this expression can be subtracted by
renormalizing the cosmological constant term.
It is because  we can attribute it to the fermion mass term.
\footnote{We recall that the minimally coupled
fermionic kinetic term on $S^2$ is $\sigma\cdot L +1$.}
The first term is the remaining conformal anomaly (geometric
entropy).
With the identification $A=(2l)^2$ and $c=8n^2$,
we confirm that the one loop correction in NC gauge theory
on $S^2$ can be understood from 2d quantum gravity point of view.

In the case of matrix quantum mechanics,
the situation is different since we have the time like
dimension in addition. Although we have analogous difficulties to
construct a model with time independent SUSY,
NC gauge theories on fuzzy spheres with
time dependent SUSY are constructed
in the context of M-theory on a pp-wave in \cite{BMN}.
The fully supersymmetric solutions of the model are
fuzzy spheres represented by $N$ dimensional
(reducible) representations of $SU(2)$.
Since the Hamiltonian can be related to the
squares of the supercharges, the vacuum energy
vanishes for all such configurations\cite{DSR}.

Although we are mostly content to compute the constant
term of the effective action in this paper,
it also depends on the zero modes.
At the one loop level, we can estimate
the effective action for bosonic
zero-modes $\bar{a}$ by putting $p_{\mu}=j_{\mu}+\bar{a}_{\mu}$
in (\ref{1lpact}).
The leading term is
\beqa
&&-2Tr[{1\over P^2}P_{\alpha}{1\over P^2}P_{\alpha}]
-2iTr[{1\over P^2}[P_{\alpha},P_{\beta}]{1\over P^2}
\epsilon^{\alpha\beta\gamma}P_{\gamma}]\n
&&= 4n^2log(2l+1) + 4 tr\bar{a}^{\alpha}\bar{a}^{\alpha}
-8tr\bar{\phi}_i\bar{\phi}_i
+\cdots .
\eeqa
We find that the coincident $n$ spheres are unstable since $\bar{\phi}_i$
become tachyonic at this order.
When two fuzzy spheres are separated by a large
distance $\phi$, there is a logarithmic repulsion between the two
\beq
8log({2l\over \phi}) .
\eeq
Such an instability is absent in 3d model of (\ref{isoact})
since there are no $\phi_i$ fields.

\section{Two loop effective action}
\setcounter{equation}{0}

In this section we summarize our two loop effective actions
and discuss their physical implications.
We delegate the detailed evaluation of
them to Appendix A.

In the case of $U(1)$ gauge group, the two loop effective action is
\beqa
F(l)&=&-36{1\over f^4}(F_3^p(l)-F_3^{np}(l)) .
\eeqa
In the case of 3d model, we obtain
\beqa
F_{3d}(l)&=&-{1\over f^4}(F_3^p(l)-F_3^{np}(l)) ,
\eeqa
where $F_3^p(l)$ and $F_3^{np}(l)$ are defined in (\ref{F3}).
We have  found numerically that
\beqa
&&F_3^p(l)\sim F_3^{np}(l)\sim 2.0{1\over N} ,\n
&&F_3^p(l) - F_3^{np}(l)\sim 6.2 {1\over N^2} .
\eeqa
Therefore we find
\beq
F_{3d}(l)=-{6.2\over f^4N^2}
=-1.5~ {g_{NC}^2\over 2\pi} .
\eeq
Since it is O(1) for fixed $g_{NC}^2$, it is much smaller than the tree
$O(N)$ or one loop $O(log(N))$ contributions.
It is because the amplitude is convergent in our model due to SUSY and
the theory becomes free (ordinary $U(1)$ gauge theory with adjoint matter)
in the infrared limit.

The situation is different for $U(n)$ gauge group.
For $U(n)$ gauge group, we obtain
\beq
F(l)=36{1\over f^4}\left(n^3 (-F_3^p(l)
+{5\over 4}F_5(l))
-n (-F_3^{np}(l)+{5\over 4}F_5(l))\right) ,
\eeq
where $F_5(l)$ is defined in (\ref{F5}).
We have numerically found that
\beq
F(l)\sim-27~{1\over f^4 (2l+1)}n(n^2-1)  +O({1/ N^2}).
\eeq
The total effective action up to the two loop
to the leading order of $1/N$ is
\beq
-{f^4\over 6}nl(l+1)(2l+1)
+4n^2log(2l+1)
-27~{1\over f^4 (2l+1)}n(n^2-1) .
\eeq
If we put $f^4=2\pi n/l^2\lambda^2$, the effective action
becomes
\beq
n^2V\left( -{2\pi \over 6\lambda^2}+4log(V)/V
-{27\over 8\pi} \lambda^2 (1-1/n^2)\right) .
\eeq
In this form, we can see that the effective action is proportional
to the volume $V=2l+1$ and $n^2$.
It also possesses the standard $1/n$ expansion where
$\lambda^2=ng_{NC}^2$ is the 't Hooft coupling.

However in a finite matrix model, we may fix $N$ and $f$ not
$V=N/n$ and $\lambda^2$. In this case a gauge group $U(n)$ is
dynamically determined by minimizing the effective action. So we need to
minimize the following expression with respect to $n$
\beq
-{f^4\over 24n^2}N^3
+4n^2log(N/n)
-27~{1\over f^4 N}n^2(n^2-1) .
\eeq
For $U(n)$ gauge group, the tree and the two loop contributions are
comparable when $f^4 N^2\sim g_{NC}^2\sim 1$.
Although the tree action is minimized when $n=1$, we observe that the two
loop contribution could easily overwhelm it if we increase $n$.
Therefore we conclude that the classically favorable
NC $U(1)$ gauge theory on fuzzy sphere is unstable at the two loop level
in a deformed IIB matrix model with a Myers term.
Since this model contains $8$ transverse degrees of freedom,
this instability may correspond to the $c=1$ barrier in 2d quantum gravity.

In the case of the 3d model, the corresponding expression is
\beq
F_{3d}(l)={1\over f^4}\left(n^3 (-F_3^p(l)
+3F_5(l))
-n (-F_3^{np}(l)+3F_5(l))\right) .
\eeq
We find numerically
\beq
F_{3d}(l)\sim{1\over f^4 (2l+1)}n(n^2-1)  +O({1/ N^2}) .
\eeq
The total effective action up to the two loop
to the leading order of $1/N$ is
\beq
-{f^4\over 6}nl(l+1)(2l+1)
+{1\over 2}n^2log(2l+1)
+{1\over f^4 (2l+1)}n(n^2-1) .
\eeq
We find again that the two loop correction is $O(N)$ for finite
$g_{NC}^2$ which is the same order with the tree action.
As it is recalled shortly, the situation is very
different  from gauge theory in flat space where SUSY cancellations take
place irrespectively of the gauge group. Since SUSY transformation
of $\psi$ field vanishes classically for any gauge group $U(n)$,
the existence of quantum corrections signals SUSY breaking effects on
$S^2$. However the two loop contributions for $U(n)$ gauge groups are
positive in sign and there is no instability of $U(1)$ gauge theory in this
model. The stability of the model is consistent with 2d quantum gravity
since it naively corresponds to $c=1$ supergravity.

\subsection*{commutative sphere limit}

In this subsection, we show that
NC gauge theory on fuzzy sphere reduces to ordinary gauge theory
on $S^2$ in the formal semiclassical (or infrared) limit.
This correspondence implies
\beq
Tr \rightarrow V\int {d\Omega\over 4\pi}tr ,
\eeq
where $d\Omega=dcos(\theta )d\varphi$.
$V=N/n$ is the volume of $S^2$  and $tr$ is over $U(n)$ gauge group.
The action (\ref{expact})
gives the following result in the semiclassical limit:
\beqa
&&{Vf^4\over 4\pi} \int{d\Omega}tr
\left( -{1\over 4}
(L_{\alpha} a_{\beta}-L_{\beta}
a_{\alpha}+[a_{\alpha},a_{\beta}])^2
-{1\over
2}(L_{\alpha}\phi_i+[a_{\alpha},\phi_i])^2\right. \n
&&
-{1\over 4}[\phi_i,\phi_j]^2
+{i\over 2}\epsilon_{\alpha\beta\gamma}
(L_{\alpha} a_{\beta}-L_{\beta} a_{\alpha})a_{\gamma}
+{i\over 3}\epsilon_{\alpha\beta\gamma}
[a_{\alpha}, a_{\beta}]a_{\gamma}  \n
&&\left .+
{1\over 2}
\bar{\psi}\gamma^{\alpha}(L_{\alpha}\psi+[a_{\alpha},\psi])+
{1\over 2}\bar{\psi}\gamma^{i}[\phi_i,\psi ]
\right) .
\eeqa
They are nothing but the ordinary gauge theory action on $S^2$.
The gauge coupling $g_{CM}^2$ is identified with
\beq
g_{CM}^2={ 2\pi \over lf^4}=lg_{NC}^2 .
\eeq
It is larger than $g_{NC}^2$ in (\ref{2dcoup}) by a factor
$l$. It is because the gauge coupling is scale dependent in
2 dimension. The factor $l$ can be understood if we recall
that the radius of the sphere is $O(\sqrt{l})$ with respect to the
non-commutativity scale and $g_{CM}^2$ is the coupling
constant at the infrared cut-off scale.

This correspondence can be proven at the
Feynman rules level.
In the perturbative investigation of ordinary gauge theory on $S^2$,
we expand the fields in terms of the spherical harmonics
\beq
\phi={1\over \sqrt{V}}\sum_{jm}\phi_{jm}Y_{jm}(\theta,\varphi ) ,
\eeq
where
\beq
\int{d\Omega\over 4\pi}Y_{j_1m_1}(\theta,\varphi )Y_{j_2m_2}(\theta,\varphi )
=\delta_{j_1,j_2}\delta_{m_1,-m_2}(-1)^{m_1} .
\eeq
The propagators are
\beq
<\phi_{j_1m_1} \phi_{j_2m_2}>={1\over f^4}{1\over
j_1(j_1+1)}\delta_{j_1,j_2}
\delta_{m_1,-m_2}(-1)^{m_1} .
\eeq
The three point vertices are given in terms of
$3j$ symbols since
\beqa
&&{f^4 \over 4\pi\sqrt{V}}\int{d\Omega}Y_{j_1m_1}(\theta,\varphi
)Y_{j_2m_2}(\theta,\varphi) Y_{j_3m_3}(\theta,\varphi )\n
&=&{f^4}\sqrt{(2j_1+1)(2j_2+1)(2j_3+1)}\n
&&\times\left(\begin{array}{ccc}
j_1&j_2&j_3\\
m_1&m_2&m_3
\end{array}\right){1 \over \sqrt{V}}
\left(\begin{array}{ccc}
j_1&j_2&j_3\\
0&0&0
\end{array}\right) .
\label{cmvrtx}
\eeqa

The Feynman rules of NC gauge theory on fuzzy sphere can be found
in Appendix A.
We can see that the propagators are identical
in the both cases as it is well known.
On the other hand,
the interaction vertices of NC gauge theory contain $6j$
symbols. In the large $l$ limit, they can be related to 3j symbols:
\beqa
\left\{\begin{array}{ccc}
j_1&j_2&j_3\\
l&l&l
\end{array}\right\}
\rightarrow {1\over \sqrt{2l}}
\left(\begin{array}{ccc}
j_1&j_2&j_3\\
0&0&0
\end{array}\right) ,
\eeqa
with $j_1,j_2,j_3$ being fixed.
Hence the interaction vertices of NC gauge theory
(\ref{6jsym}) reduces to
those of ordinary gauge theory expression (\ref{cmvrtx})
as $V\sim 2l$.
We can thus observe that the former reduces to the latter
in the formal semi-classical limit at the Feynman rules level.

\subsection*{U(1) Yang-Mills theory on non-commutative
${ R^2}$}
In this section, we have investigated the 2-loop effective action of
the matrix models on fuzzy sphere.
In the remainder we compare the results with
the 2-loop 1PI contribution to vacuum energy in
U(1) Yang-Mills theory on fuzzy ${ R^2}$.

We quote the action:
\begin{eqnarray}
S&=&{1\over g_{NC}^2}\int d^2x\;{\cal L},\n
{\cal L}&=&-\frac{1}{2}A^{\mu}\partial^2 A_{\mu}
-b\partial^2\; c
-\frac{i}{2}tr\bar{\Psi}\Gamma^{\mu}\partial_{\mu}\Psi\n
&&+i\partial_{\mu}b*[c,A^{\mu}]_{*}
-\frac{1}{2}tr\bar{\Psi}*\Gamma^{\mu}[A_{\mu},\Psi]_{*}\n
&&-\frac{i}{2}(\partial_{\mu}A_{\nu}-\partial_{\nu}A_{\mu})*[A^{\mu},
A^{\nu}]_{*}
-\frac{1}{4}[A_{\mu},A_{\nu}]_{*}[A^{\mu},A^{\nu}]_{*}  .
\end{eqnarray}
One can obtain this action locally from the matrix models
as it is explained in the preceding section
after the identification of $g_{NC}^2=2\pi/f^4l^2$.
In this theory vacuum energy contributions from the 2loop 1PI graphs
can be evaluated as follows:

\begin{itemize}
\item
2-loop 1PI from 4-gauge boson vertex:
\begin{eqnarray}
-45g_{NC}^2V\cdot \int {d^2p\over (2\pi)^2} \int
{d^2q \over (2\pi)^2}\;\frac{1}{p^2q^2}
\;(1-\cos(p\wedge q)) ,
\end{eqnarray}
where $V$ denotes the volume of ${ R^2}$.

\item
2-loop 1PI from 3-gauge boson vertices:
\begin{eqnarray}
9g_{NC}^2V\cdot \int {d^2p\over (2\pi)^2} \int
{d^2q \over (2\pi)^2}\;
\left(\frac{1}{p^2q^2}-\frac{p\cdot q}{p^2q^2(p+q)^2}\right)
\;(1-\cos(p\wedge q)) .
\end{eqnarray}

\item
2-loop 1PI from ghost-gauge vertices:
\begin{eqnarray}
g_{NC}^2V\cdot\int {d^2p\over (2\pi)^2} \int
{d^2q \over (2\pi)^2}\;
\frac{p\cdot q}{p^2q^2(p+q)^2}
\;(1-\cos(p\wedge q)) .
\end{eqnarray}

\item
2-loop 1PI from fermion-gauge vertices:
\begin{eqnarray}
-64g_{NC}^2V\cdot\int {d^2p\over (2\pi)^2} \int
{d^2q \over (2\pi)^2}\;
\frac{p\cdot q}{p^2q^2(p+q)^2}
\;(1-\cos(p\wedge q)).
\end{eqnarray}
\end{itemize}

The total amplitude is
\begin{eqnarray}
&&-36g_{NC}^2V\cdot\int {d^2p\over (2\pi)^2} \int
{d^2q \over (2\pi)^2}\;
\frac{1}{p^2q^2}
\;(1-\cos(p\wedge q)) \n
&&-72g_{NC}^2V\cdot\int {d^2p\over (2\pi)^2} \int
{d^2q \over (2\pi)^2}\;
\frac{p\cdot q}{p^2q^2(p+q)^2}
\;(1-\cos(p\wedge q)) .
\end{eqnarray}
The first and the second term cancel each other
since
\beq
\frac{p\cdot q}{p^2q^2(p+q)^2}=
\frac{1}{2}\left(\frac{1}{p^2q^2}-\frac{1}{q^2(p+q)^2}
-\frac{1}{p^2(p+q)^2}\right) .
\eeq

By comparing these amplitudes with
those in Appendix A, we can find the following correspondence
between the amplitudes of
NC $U(1)$ gauge theory on ${ R^2}$ and $S^2$:
\begin{eqnarray}
&&\int {d^2p\over (2\pi)} \int
{d^2q \over (2\pi)}\;
\frac{1}{p^2q^2}
\;(1-\cos(p\wedge q)) \quad \Leftrightarrow \quad
{l\over 2}(F_1^p\left(l\right)-F_1^{np}\left(l\right)) ,\n
&&\int {d^2p\over (2\pi)} \int
{d^2q \over (2\pi)}\;
\frac{p\cdot q}{p^2q^2(p+q)^2}
\;(1-\cos(p\wedge q))
\quad \Leftrightarrow\quad
{l\over 2}
\left(F_2^p\left(l\right)-F_2^{np}\left(l\right)\right) ,
\end{eqnarray}
where $F_1^{p(np)}(l)$ and $F_2^{p(np)}(l)$ are defined in (\ref{F1})

Non-planar phases: $\cos(p\wedge q)$
in NC gauge theories on ${ R^2}$ correspond to
$(-1)^{(j_1+ \cdots)}$ in those on $S^2$.
We point out here that the
$6j$ symbols vanish unless the non-planar
phases are trivial in the semiclassical (infrared) limit.
Therefore the non-commutativity plays no role
in the infrared limit in the both theories.
We further observe the identical power counting rules
in the both theories. Namely logarithmically
divergent amplitudes and convergent amplitudes
correspond to each other.

The amplitudes of NC gauge theory on fuzzy $S^2$, however, are different
from  those on ${ R^2}$ in the following points:
\begin{itemize}
\item
In NC gauge theory on fuzzy ${ R^2}$,
the cancellation of 2-loop effective action occurs in the planer and
non-planar sectors separately.  In NC gauge theory on fuzzy $S^2$, the
corresponding cancellation does not occur if
we consider planar or non-planar contributions separately.

\item
In NC gauge theory on fuzzy $S^2$, the fermionic contribution gives
rise to extra terms represented by $F_3^p$ and $F_3^{np}$.
They result in the non-vanishing
effective action even for $U(1)$ gauge theory.
The origin of this new term may be thought as
the spin connection of fermions in a curved space.
\end{itemize}

\section{Conclusions and Discussions}
\setcounter{equation}{0}
In this paper, we have investigated quantum corrections
in NC gauge theory on fuzzy sphere up to the two loop level.
Such theories are realized by matrix models with a Myers term.
The classical solutions of the matrix models are
reducible representations of $SU(2)$ which represent
a group of fuzzy spheres. The irreducible representation which represents
a single fuzzy sphere minimizes the classical action.
These backgrounds are supersymmetric in the sense that
the SUSY transformation of $\psi$ field classically vanishes.

We have found that the quantum corrections do not vanish at each order.
At the one loop level, it does not vanish due to the presence of
the cut-off for the angular momentum. Due to the mismatch between
the bosonic and fermionic degrees of freedom at the largest
angular momentum, we find $O(log(N))$ one loop contribution.
At the two loop level, we find comparable quantum corrections
to the tree action with finite gauge coupling $g_{NC}$ in the case
of $U(n)$ gauges group. It is because SUSY cancellations do
not take place in planar and non-planar sectors separately
contrary to flat theory. In a deformed IIB matrix model with
a Myers term, we find this effect destabilize the classically
favorable $U(1)$ gauge theory.

Our results are consistent with the fact that
there cannot be unbroken supersymmetry in de-Sitter space
since the Euclidean continuation of an $n$-dimensional de-Sitter space
$dS_n$ is an $n$ sphere $S^n$.
We may put forward a generic argument as follows.
In any field theory on a homogeneous space $G/H$,
we need to identify the Hamiltonian with one of the Killing
vectors which form the Lie algebra of $G$.
Since there is no positive Killing vectors
in this case, there is no positive conserved energy.
It in turn implies that there is no supercharge $Q$
whose square equals the Hamiltonian.
Consequently there cannot be unbroken supersymmetry
in compact homogeneous spaces.

One of the most important questions concerning NC gauge theory
is its relation to quantum gravity.
Various similarities have been pointed out
such as the absence of local gauge invariant operators,
UV-IR mixing and Wilson lines which couple to closed strings.
In this paper we have found that NC gauge theory on $S^2$
allows 2d quantum gravitational interpretation.
We have reproduced the semiclassical free energy of
2d supergravity. We have further seen a possible signal for
$c=1$ barrier. It is therefore tempting to conjecture that
NC gauge theory on $S^2$ belongs to the same universality class
of 2d quantum gravity.

Since we can locally recover flat theory with SUSY
in these models,
the SUSY breaking effect must come from those
degrees of freedom which is sensitive
to the curvature. They are either infra-red (IR)
or ultra-violet (UV) degrees of freedom.
Since the two loop amplitudes are finite due to SUSY
in our case, we are mostly sensitive to IR degrees of freedom.
In the case of $U(1)$ gauge theory, we find little quantum
corrections since it becomes free theory in IR regime.

In the case of higher dimensional spaces like
$CP^2$ or $S^2\times S^2$,
the effective actions are no longer convergent in the
sense that they are dominated by IR contributions.
In that case,  we expect that
UV degrees of freedom also contribute to the
SUSY breaking effect.
It will be interesting to generalize
our investigations to higher dimensional homogeneous
spaces.
It is also interesting to explore the relevance
of these spaces to IIB matrix model
in the spirit of \cite{NS}\cite{Kyoto}.

\begin{center} \begin{large}
Acknowledgments
\end{large} \end{center}
We would like to thank M. Hatsuda, S. Iso and T. Suyama
for discussions.
This work is supported in part by the Grant-in-Aid for Scientific
Research from the Ministry of Education, Science and Culture of Japan.

\section*{Appendix A}
\renewcommand{\theequation}{A.\arabic{equation}}

In this appendix, we evaluate
the two loop effective action
of NC gauge theory on fuzzy sphere.
We consider NC gauge theory with $U(1)$ gauge group in the context of
a deformed IIB matrix model with a Myers term.
We also explain how to modify our results in the case of the 3d model
or different gauge groups.

There are 5 diagrams to evaluate which are
illustrated in  Figure 1.
(a),(b) and (c) represent contributions from gauge fields.
(a) and (b) are of different topology while (c)
involves the Myers type interaction.
(d) involves ghost and (e) fermions respectively.

\begin{figure}[hbtp]
\epsfysize=3cm
\begin{center}
\vspace{1cm}
\epsfbox{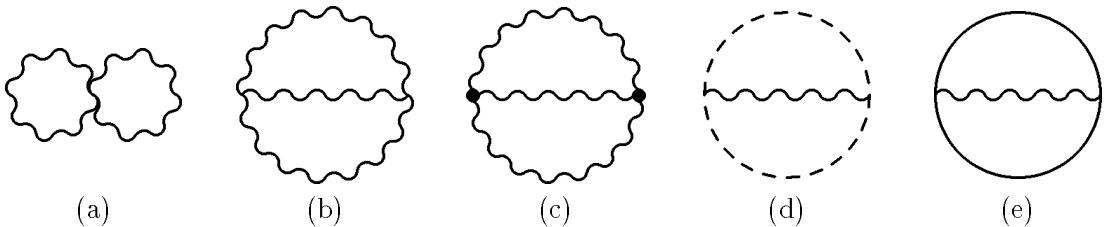}
\end{center}
\caption{Feynman diagrams of 2 Loop corrections to the effective action}
\label{Fig:2LoopCorrections}
\end{figure}

We expand matrices in terms of matrix spherical harmonics:
\beqa
&&A_{\mu}=p_{\mu}+\sum_{jm}a^{\mu}_{jm}Y_{jm} ,\n
&&\psi=\sum_{jm}\psi_{jm}Y_{jm} ,
\eeqa
where $p_{\alpha}=j_{\alpha}$  and other $p_{\mu}'s=0$.
We adopt the following representation of $Y_{jm}$:
\begin{eqnarray}
&&(Y_{jm})_{ss'}=(-1)^{l-s}
\left(\begin{array}{ccc}
l&j&l_{} \\
-s&m&s'
\end{array}\right)
\sqrt{2j+1} .
\end{eqnarray}
where they are normalized as
\beq
\mbox{Tr}\;Y_{j_1m_1}Y_{j_2m_2}=(-1)^{m_1}\delta_{j_1,j_1}
\delta_{m_1,-m_{2}} .
\eeq
The cubic couplings of the matrix spherical harmonics
can be evaluated as
\beqa
&&Tr[Y_{j_1m_1}Y_{j_2m_2}Y_{j_3m_3}]\n
&&=(-1)^{2l}\sqrt{(2j_1+1)(2j_2+1)(2j_3+1)}\n
&&\times\left(\begin{array}{ccc}
j_1&j_2&j_3\\
m_1&m_2&m_3
\end{array}\right)
\left\{\begin{array}{ccc}
j_1&j_2&j_3\\
l&l&l
\end{array}\right\} .
\label{6jsym}
\eeqa
We refer to \cite{Edm} for $( 3j   )$ and  $\{6j \}$ symbols.

\subsubsection*{bosonic propagators }
From the quadratic terms in the gauge fixed action, we can read propagators
of gauge boson modes $a^{\mu}_{jm}$ and ghost modes $b_{jm}$, $c_{jm}$ as
follows
\begin{eqnarray}
\langle\; a^{\mu}_{j_1m_1}a^{\nu}_{j_2m_2}\;\rangle&=&
\frac{1}{f^{4}}\;\frac{(-1)^{m_1}}{j_1(j_1+1)}\;
\delta^{\mu\nu}\delta_{j_1j_2}\delta_{m_1-m_{2}} ,
\n
\langle\; c_{j_1m_1}b_{j_2m_2}\;\rangle&=&
\frac{1}{f^{4}}\;
\frac{(-1)^{m_1}}{j_1(j_1+1)}\;\delta_{j_1j_2}\delta_{m_1-m_{2}} .
\end{eqnarray}
In terms of fields
\begin{eqnarray}
a^{\mu}_{st}&=&\sum_{jm}\;a^{\mu}_{jm}(Y_{jm})_{st}, \n
b_{st}&=&\sum_{jm}\;b_{jm}(Y_{jm})_{st}, \n
c_{st}&=&\sum_{jm}\;c_{jm}(Y_{jm})_{st},
\end{eqnarray}
propagators become
\begin{eqnarray}
\langle\; a^{\mu}_{st}\;a^{\nu}_{uv}\;\rangle&=&
\frac{1}{f^{4}}\;
\sum_{jm}\;\frac{(-1)^{m}}{j_1(j_1+1)}\;\delta^{\mu\nu}
(Y_{jm})_{st}(Y_{j-m})_{uv} ,
\n
\langle\; c_{st}\;b_{uv}\;\rangle&=&
\frac{1}{f^{4}}\;\sum_{jm}\;
\frac{(-1)^{m}}{j_1(j_1+1)}\;
(Y_{jm})_{st}(Y_{j-m})_{uv} .
\end{eqnarray}

\subsection*{contribution from 3-gauge boson vertices (b)}
Firstly we calculate 2-loop 1PI contribution from 3-gauge boson vertices:
\begin{eqnarray}
V_3=\frac{f^{4}}{2}{\mbox Tr}\;[p_{\mu},a_{\nu}][a_{\mu},a_{\nu}]
-\frac{f^{4}}{2}{\mbox Tr}\;[p_{\nu},a_{\mu}][a_{\mu},a_{\nu}].
\end{eqnarray}
There are $3!=6$ type contractions following Wick's theorem
in the calculation of $\langle\;{1\over 2}V_3V_3\;\rangle$.
The result is that:
 \begin{eqnarray}
&&\frac{10-1}{2}\;\frac{1}{f^{4}}
\sum_{j_1j_2j_3=1}^{2l}\sum_{m_1m_2m_3}
\frac{(-1)^{m_1+m_2+m_3}}{j_1(j_1+1)j_2(j_2+1)j_3(j_3+1)}\;\n
&&\;\times\;
\mbox{Tr}\left([p_{\mu}, Y_{j_1m_1}][Y_{j_2m_2},Y_{j_3m_3}]\right)
\n
&&
\times\Bigg(
\mbox{Tr}\left([p_{\mu}, Y_{j_1-m_1}][Y_{j_2-m_2},Y_{j_3-m_3}]\right)
+\mbox{Tr}\left([p_{\mu}, Y_{j_2-m_2}][Y_{j_1-m_1},Y_{j_3-m_3}]\right)
\Bigg).
\label{3vertx}
\end{eqnarray}

\subsubsection*{3j-,6j- symbol expression of the results}
We can further evaluate the matrix representation (\ref{3vertx}) in terms
of $3j$ and $6j$ symbols.
The adjoint $P_{\alpha}\equiv[p_{\alpha},\;\;]$ act on $Y_{jm}$ as
\begin{eqnarray}
[\;p_{+},\; Y_{jm}\;]&=&\sqrt{(j-m)(j+m+1)} \;Y_{jm+1} ,
\qquad p_{+}=p_1+ip_2 ,  \n
{}[\;p_{-}, \;Y_{jm}\;]&=&\sqrt{(j+m)(j-m+1)} \;Y_{jm-1} ,
\qquad p_{-}=p_1-ip_2 ,  \n
{}[\;p_3, \;Y_{jm}\;]&=&m \;Y_{jm} .
\end{eqnarray}
Using this formula and the expression:
\begin{eqnarray}
&&\mbox{Tr}\left(Y_{j_1m_1}[Y_{j_2m_2},Y_{j_3m_3}]\right)
\n&&=(1-(-1)^{j_1+j_2+j_3})(-1)^{2l}
\sqrt{(2j_1+1)(2j_2+1)(2j_3+1)}
\left(\begin{array}{ccc}
j_1&j_2&j_3\\
m_1&m_2&m_3
\end{array}\right)
\left\{\begin{array}{ccc}
j_1&j_2&j_3\\
l&l&l
\end{array}\right\},\n
\end{eqnarray}
one can obtain
\begin{eqnarray}
&&9\cdot\frac{1}{f^{4}}
\sum_{j_1j_2j_3=1}^{2l}\sum_{m_1m_2m_3}(1-(-1)^{j_1+j_2+j_3})
\frac{(2j_1+1)(2j_2+1)(2j_3+1)}{j_2(j_2+1)j_3(j_3+1)}\;\n
&&\times
\left(\begin{array}{ccc}
j_1&j_2&j_3\\
m_1&m_2&m_3
\end{array}\right)^2
\left\{\begin{array}{ccc}
j_1&j_2&j_3\\
l&l&l
\end{array}\right\}^2\n
&&+9\cdot\frac{1}{f^{4}}
\sum_{j_1j_2j_3=1}^{2l}\sum_{m_1m_2m_3}(1-(-1)^{j_1+j_2+j_3})
\frac{(2j_1+1)(2j_2+1)(2j_3+1)}{j_1(j_1+1)j_2(j_2+1)j_3(j_3+1)}
\left\{\begin{array}{ccc}
j_1&j_2&j_3\\
l&l&l
\end{array}\right\}^2\n
&&\times \Bigg[
-\frac{1}{2}\sqrt{(j_1-m_1)(j_1+m_1+1)(j_2+m_2)(j_2-m_2+1)}\n
&&\times
\left(\begin{array}{ccc}
j_1&j_2&j_3\\
m_1+1&m_2&m_3
\end{array}\right)
\left(\begin{array}{ccc}
j_2&j_1&j_3\\
-m_2-1&-m_1&-m_3
\end{array}\right)\n
&&-\frac{1}{2}\sqrt{(j_1+m_1)(j_1-m_1+1)(j_2-m_2)(j_2+m_2+1)}\n
&&\times
\left(\begin{array}{ccc}
j_1&j_2&j_3\\
m_1-1&m_2&m_3
\end{array}\right)
\left(\begin{array}{ccc}
j_2&j_1&j_3\\
-m_2+1&-m_1&-m_3
\end{array}\right)\n
&&-m_1m_2
\left(\begin{array}{ccc}
j_1&j_2&j_3\\
m_1&m_2&m_3
\end{array}\right)\left(\begin{array}{ccc}
j_2&j_1&j_3\\
-m_2&-m_1&-m_3
\end{array}\right)\Bigg] .
\end{eqnarray}

In the first line, one can use the property of 3j-symbol:
\begin{eqnarray}
\sum_{m_3}\sum_{m_1m_2}
\left(\begin{array}{ccc}
j_1&j_2&j_3\\
m_1&m_2&m_3
\end{array}\right)^2
=\sum_{m_3}\frac{1}{2j_3+1}=1 ,
\end{eqnarray}
and the property of 6j-symbol:
\begin{eqnarray}
\sum_{j_1=1}^{2l}(2j_1+1)
\left\{\begin{array}{ccc}
j_1&j_2&j_3\\
l&l&l
\end{array}\right\}^2
&=&\frac{1}{2l+1}-\frac{1}{(2l+1)(2j_2+1)}\delta_{j2,j3}
,\n
\sum_{j_1=1}^{2l}\;(-1)^{j_1}(2j_1+1)
\left\{\begin{array}{ccc}
j_1&j_2&j_3\\
l&l&l
\end{array}\right\}^2
&=&\left\{\begin{array}{ccc}
l&l&j_2\\
l&l&j_3
\end{array}\right\}
-\frac{1}{(2l+1)(2j_2+1)}\delta_{j2,j3} .
\label{singlet}
\end{eqnarray}
Here we need to recall that $SU(2)$ singlet states are absent
in the gluon propagator which results in the extra terms on the
right-hand side of (\ref{singlet}). Although they cancel
each other for $U(1)$ case, it is not the case for $U(n)$
gauge groups.

In this way one can rewrite (\ref{3vertx}) as
\beq
\frac{9}{f^{4}}
\Bigg(F_1^p(l)-F_1^{np}(l)
-F_2^p(l)+F_2^{np}(l)\Bigg) ,
\eeq
where
\beqa
&&F_1^p\left(l\right)=\frac{1}{2l+1}\sum_{j_1j_2=1}^{2l}
\frac{(2j_1+1)(2j_2+1)}{j_1(j_1+1)j_2(j_2+1)} ,\n
\n
&&F_1^{np}\left(l\right)=\sum_{j_1j_2=1}^{2l}(-1)^{j_1+j_2}
\frac{(2j_1+1)(2j_2+1)}{j_1(j_1+1)j_2(j_2+1)}
\left\{\begin{array}{ccc}
l&l&j_2\\
l&l&j_1
\end{array}\right\} ,\n
\n
&&F_2^p\left(l\right)=
\sum_{j_1j_2j_3=1}^{2l}\sum_{m_1m_2m_3}
\frac{(2j_1+1)(2j_2+1)(2j_3+1)}{j_1(j_1+1)j_2(j_2+1)j_3(j_3+1)}
\left\{\begin{array}{ccc}
j_1&j_2&j_3\\
l&l&l
\end{array}\right\}^2
\n
&&\hspace{-10mm}\times \Bigg[
\frac{1}{2}\sqrt{(j_1-m_1)(j_1+m_1+1)(j_2+m_2)(j_2-m_2+1)}
\left(\begin{array}{ccc}
j_1&j_2&j_3\\
m_1+1&m_2&m_3
\end{array}\right)
\left(\begin{array}{ccc}
j_1&j_2&j_3\\
m_1&m_2+1&m_3
\end{array}\right)
\n
&&+\frac{1}{2}\sqrt{(j_1+m_1)(j_1-m_1+1)(j_2-m_2)(j_2+m_2+1)}
\left(\begin{array}{ccc}
j_1&j_2&j_3\\
m_1-1&m_2&m_3
\end{array}\right)
\left(\begin{array}{ccc}
j_1&j_2&j_3\\
m_1&m_2-1&m_3
\end{array}\right)
\n
&&\hspace{40mm}+m_1m_2
\left(\begin{array}{ccc}
j_1&j_2&j_3\\
m_1&m_2&m_3
\end{array}\right)^2\;\Bigg] ,\n
\n
&&F_2^{np}\left(l\right)=
\sum_{j_1j_2j_3=1}^{2l}\sum_{m_1m_2m_3}(-1)^{j_1+j_2+j_3}
\frac{(2j_1+1)(2j_2+1)(2j_3+1)}{j_1(j_1+1)j_2(j_2+1)j_3(j_3+1)}
\left\{\begin{array}{ccc}
j_1&j_2&j_3\\
l&l&l
\end{array}\right\}^2
\n
&&\hspace{-10mm}\times \Bigg[
\frac{1}{2}\sqrt{(j_1-m_1)(j_1+m_1+1)(j_2+m_2)(j_2-m_2+1)}
\left(\begin{array}{ccc}
j_1&j_2&j_3\\
m_1+1&m_2&m_3
\end{array}\right)
\left(\begin{array}{ccc}
j_1&j_2&j_3\\
m_1&m_2+1&m_3
\end{array}\right)
\n
&&+\frac{1}{2}\sqrt{(j_1+m_1)(j_1-m_1+1)(j_2-m_2)(j_2+m_2+1)}
\left(\begin{array}{ccc}
j_1&j_2&j_3\\
m_1-1&m_2&m_3
\end{array}\right)
\left(\begin{array}{ccc}
j_1&j_2&j_3\\
m_1&m_2-1&m_3
\end{array}\right)
\n
&&\hspace{40mm}+m_1m_2
\left(\begin{array}{ccc}
j_1&j_2&j_3\\
m_1&m_2&m_3
\end{array}\right)^2\;\Bigg] .
\label{F1}
\eeqa

The overall coefficient $9=10-1$ changes to $2$ for the 3d model
since $2=3-1$.

\subsection*{ghost contribution (d)}
We secondly calculate the contribution from the ghost-gauge boson vertex:
\begin{eqnarray}
V_{gh}=-f^{4}\mbox{Tr}[p_{\mu}, b][c,a_{\mu}] .
\end{eqnarray}
The result is that:
\begin{eqnarray}
&&-\frac{1}{2}\frac{1}{f^{4}}\sum_{j_1j_2j_3}\sum_{m_1m_2m_3}\;
\frac{(-1)^{m_1+m_2+m_3}}{j_1(j_1+1)j_2(j_2+1)j_3(j_3+1)}\;\n
&&\qquad
\times\quad \mbox{Tr}\big([p_{\mu},Y_{j_1m_1}][Y_{j_2m_2},
Y_{j_3m_3}]\big)\cdot
\mbox{Tr}\big([p_{\mu},Y_{j_2-m_2}][Y_{j_1-m_1}, Y_{j_3-m_3}]\big).
\end{eqnarray}
It has the same form as
the 2nd term in (\ref{3vertx}).

\subsection*{contribution from 4-gauge boson vertex (a)}

The interaction vertex which involves 4 gauge bosons is
\begin{equation}
V_4 = - \frac{f^4}{4} \textrm{Tr} \, [ a_{\mu},a_{\nu} ] [
a_{\mu},a_{\nu} ] .
\end{equation}
It gives rises to the following contribution
\begin{eqnarray}
&&< - \frac{1}{1!} V_4 >_{\textrm{1PI-2loop}} \n
& = &  \frac{1}{4} (10^2-10)
\frac{1}{f^4}  \sum_{j_1 j_2 =1}^{2l} \sum_{m_1 m_2 }
\frac{(-)^{m_1 + m_2} \ \textrm{Tr} \, [ Y_{j_1 m_1} ,Y_{j_2 m_2} ]
[ Y_{j_1 -m_1} ,Y_{j_2 -m_2} ]}{j_1 (j_1 +1)j_2 (j_2 +1)}\n
& = & - \frac{45}{2}  \frac{1}{f^4}
\sum_{j_1 j_2 j_3=1}^{2l}
\sum_{m_1 m_2 m_3} j_1 (j_1 +1) \frac{(\textrm{Tr} \,
Y_{j_1 m_1}
 [Y_{j_2 m_2} ,Y_{j_3 m_3}] )^2}{j_1 (j_1 +1)j_2 (j_2 +1)j_3 (j_3 +1)}\n
& = & - 45
\frac{1}{f^4}
\sum_{j_1 j_2 j_3=1}^{2l} \sum_{m_1 m_2 m_3} j_1 (j_1 +1)
\frac{(2j_1 +1)(2j_2 +1)(2j_3 +1)}{j_1 (j_1 +1)j_2 (j_2 +1)j_3 (j_3 +1)} \n
& & \hspace{5mm}
\times \  (1-(-1)^{j_1 +j_2 +j_3})  \left(\begin{array}{ccc}
j_1&j_2&j_3\\
m_1&m_2&m_3
\end{array}\right)^2
\left\{\begin{array}{ccc}
j_1&j_2&j_3\\
l&l&l
\end{array}\right\}^2
\n & = & - {45}\frac{1}{f^4}
\left( F_1^p(l)-F_1^{np}(l)\right) .
\end{eqnarray}
In the 3d model, the corresponding amplitude is
\beq
- {3}\frac{1}{f^4}
\left(F_1^p(l)-F_1^{np}(l)\right) .
\eeq

\subsection*{contribution from cubic vertices (c)}
The cubic vertex which contains the structure constant of $SU(2)$ is
\begin{equation}
V_{cubic} =  \frac{i}{3} f^4 \epsilon_{\mu \nu \rho} \textrm{Tr} \,
[ a_{\mu},a_{\nu} ]
a_{\rho} .
\end{equation}
Their contribution is
\begin{eqnarray}
&&< \frac{1}{2!} V_{cubic} V_{cubic}  >_{\textrm{1PI-2loop}}\n
& = &  2 \frac{1}{f^4}  \sum_{j_1 j_2 j_3=1}^{2l} \sum_{m_1 m_2 m_3}
\frac{(\textrm{Tr} \,  Y_{j_1 m_1}  [Y_{j_2 m_2} ,Y_{j_3 m_3}] )^2}
{j_1 (j_1+1)j_2 (j_2 +1)j_3 (j_3 +1)}\n
& = & 4 \frac{1}{f^4} \sum_{j_1 j_2 j_3=1}^{2l} \sum_{m_1 m_2 m_3}
(1-(-1)^{j_1 +j_2 +j_3})
\frac{(2j_1 +1)
(2j_2 +1)(2j_3 +1)}{j_1 (j_1 +1)j_2 (j_2 +1)j_3 (j_3 +1)}
\n
&&\times
\left(\begin{array}{ccc}
j_1&j_2&j_3\\
m_1&m_2&m_3
\end{array}\right)^2
\left\{\begin{array}{ccc}
j_1&j_2&j_3\\
l&l&l
\end{array}\right\}^2
\n
& = &  \frac{4}{f^4}
(F_3^p(l)-F_3^{np}(l)) ,
\end{eqnarray}

where

\begin{eqnarray}
F_3^p\left(l\right) &=&
\sum_{j_1,j_2,j_3=1}^{2l}
     {\left(2j_1+1\right) \left(2j_2+1\right) \left(2j_3+1\right)
                          \over j_1\left(j_1+1\right)j_2\left(j_2+1\right)
j_3\left(j_3+1\right)}
     \left\{
 \begin{array}{ccc}
  j_1 &  j_2 & j_3 \\
  l   &  l & l
 \end{array}
\right\}^2 , \n
F_3^{np}\left(l\right) &=&
\sum_{j_1,j_2,j_3=1}^{2l}
\left(-1\right)^{j_1+j_2+j_3}
     {\left(2j_1+1\right) \left(2j_2+1\right) \left(2j_3+1\right)
                          \over j_1\left(j_1+1\right)j_2\left(j_2+1\right)
j_3\left(j_3+1\right)}
     \left\{
 \begin{array}{ccc}
  j_1 &  j_2 & j_3 \\
  l   &  l & l
 \end{array}
\right\}^2 .\n
\label{F3}
\end{eqnarray}

\section*{fermion propagators and vertices }

The fermionic action is
\begin{eqnarray}
S^\psi = -{f^4 \over 2} \mbox{Tr} \left( {\bar \psi}
\Gamma^\mu \left[A_\mu, \psi\right]\right),
\end{eqnarray}
where we have scaled out $f$.

Specially we can represent the $\Gamma^\mu$ matrices as follows
\begin{eqnarray}
 \Gamma^0 &=&
\left(
 \begin{array}{cc}
  0        & i 1_{16} \\
 -i 1_{16} & 0
\end{array}
\right), \quad
\Gamma^9 =
\left(
\begin{array}{cc}
  0            & {\gamma^9} \\
 {\gamma^9}^T & 0
\end{array}
\right), \quad
\Gamma^8 =
\left(
 \begin{array}{cc}
 0            & \gamma^8 \\
 {\gamma^8}^T & 0
\end{array}
\right), \n
\Gamma^i &=&
\left(
 \begin{array}{cc}
  0        & \gamma^i \\
 {\gamma^i}^T & 0
 \end{array}
\right)
\quad (i=1, 2 \cdots, 7) ,
\end{eqnarray}
where
\begin{eqnarray}
\gamma^8  &=&
\left(
\begin{array}{cc}
 0   & 1_8 \\
 1_8 & 0
\end{array}
\right), \quad
\gamma^9  =
\left(
\begin{array}{cc}
  1_8 &  0   \\
  0   & -1_8
\end{array}
\right), \n
\gamma^i &=&
\left(
\begin{array}{cc}
 0                 & {\gamma^i}_8 \\
 -{\gamma^i}_8 & 0
\end{array}
\right) \quad \left(i=1, 2 \cdots, 7 \right), \quad
\end{eqnarray}
and $\gamma^i_8$ are real and antisymmetric.

The fermion $\psi$ is written by using chiral projection
\begin{eqnarray}
\psi={1+ \Gamma \over 2}
\left(
\begin{array}{c}
\psi_{16}  \\
\psi_{16}'
\end{array}
\right), \n
\Gamma = -i \Gamma^0 \Gamma^1 \cdots \Gamma^9,
\end{eqnarray}
We can rewrite the fermion action as
\begin{eqnarray}
S^\psi = {f^4 \over 2} \mbox{Tr} \left[\psi_{16}^T \tilde{\gamma}^0
\tilde{\gamma}^{\mu} [A_{\mu},\psi_{16}] \right] .
\end{eqnarray}
The fermion kinetic term is
\beq
S_0^\psi = {f^4 \over 2} \mbox{Tr} \left[\psi_{16}^T \tilde{\gamma}^0
\tilde{\gamma}^{\alpha} [p_{\alpha},\psi_{16}] \right],
\eeq
where
\beqa
&&\tilde{\gamma}^0=\sigma^2\otimes 1_8,
\quad\tilde{\gamma}^9=-\sigma^1\otimes 1_8,
\quad\tilde{\gamma}^8=\sigma^3\otimes 1_8,\n
&&\tilde{\gamma}^i=-1_2\otimes \gamma^i_8
\quad (i=1,\cdots, 7) .
\eeqa
The interaction vertices are
\beq
S_1^\psi = {f^4 \over 2} \mbox{Tr} \left[\psi_{16}^T \tilde{\gamma}^0
\tilde{\gamma}^{\alpha} [a_{\alpha},\psi_{16}] \right]
+{f^4 \over 2} \mbox{Tr} \left[\psi_{16}^T \tilde{\gamma}^0
\tilde{\gamma}^{i} [\phi_{i},\psi_{16}] \right] .
\eeq

The kinetic term can be rewritten as
\beq
S_0^\psi = {f^4 \over 2} \mbox{Tr} \left[\psi_{a}^T \sigma^2
\sigma^{\alpha} [p_{\alpha},\psi_{a}] \right],
\label{ferkin}
\eeq
where $\psi_{a}$ are 8 SU(2) doublets.
From (\ref{ferkin}), the fermion propagator is
\begin{eqnarray}
\langle \left(\psi_{a}\right)_{\beta,st} \left(\psi
_{b}{\sigma^2}\right)_{\delta,uv}
\rangle &=& {1 \over f^4}
\sum_{j_1=1}^{2l}
\sum_{m=-\left(j_1+{1 \over 2} \right)}^{j_1+{1 \over 2}}
\left\{-{1 \over j_1+1}  C_{-}\left(j_1 \enspace \beta \enspace \delta
\enspace m
\right) +{1 \over j_1}  C_{+}\left(j_1 \enspace \beta \enspace \delta
\enspace m \right) \right\}  \nonumber \\
&& \times \left(Y_{j_1,m-\beta}^{}\right)_{st}
\left(Y_{j_1,m-\delta}^\dag\right)_{uv}
\delta_{ab} ,
\end{eqnarray}
where $s, t, u, v$ are matrix indices, $\beta,\delta$ are $SU(2)$ spin
indices and  $a, b$ are 8 `flavor' indices.
$C_{-},C_{+}$ are defined as follows:
\begin{eqnarray}
C_{+}(j_1 \enspace \pm {1 \over 2} \enspace  \pm{1 \over 2} \enspace m )
&=&
-C_{-}(j_1 \enspace \mp {1 \over 2} \enspace  \mp{1 \over 2} \enspace m )
=
{j_1 \pm m + {1 \over 2} \over 2 j_1 +1}, \n
C_{-}(j_1 \enspace \pm {1 \over 2} \enspace \mp{1 \over 2} \enspace m)
&=&
-{\sqrt{(j_1 -m + {1 \over 2})(j_1+ m + {1 \over 2})} \over 2 j_1 +1}, \n
C_{+}(j_1 \enspace \pm {1 \over 2} \enspace \mp{1 \over 2} \enspace m)
&=&
{\sqrt{(j_1 -m + {1 \over 2})(j_1+ m + {1 \over 2})} \over 2 j_1 +1}.
\end{eqnarray}

\section*{fermionic contribution (e)}

2-loop (1PI) contribution to the effective action from the fermion-boson
vertex is
\begin{eqnarray}
&&{1 \over 2!} \Big\langle
 (-S_{1}^\psi) (-S_{1}^\psi )\Big\rangle_{\mbox{2loop-1PI}}\n
&=&
{1 \over f^4}
\sum_{j_1 j_2 j_3} \sum_{m_1 m_2 m_3}\n
&&
\left(
4\cdot (7-3){\mbox{Tr}\{(Y_{j_1 m_1} [Y_{j_2 m_2},Y_{j_3 m_3}])^2\} \over
j_1(j_1+1) j_2(j_2+1) j_3(j_3+1)} \right.
\n
& &
+4\cdot (7+1){(-1)^{m_1+m_2+m_3} \over j_1 (j_1+1) j_2 (j_2+1) j_3(j_3+1)}
\n
&&\times \left.\mbox{Tr} [p_\mu, Y_{j_1m_1}][Y_{j_2m_2},Y_{j_3m_3}]
\mbox{Tr} [p_\mu,
Y_{j_2,-m_2}][Y_{j_1,-m_1},Y_{j_3,-m_3}]
\right)
\n &=&
{32 \over f^4} \left[ F_3^p\left(l\right) - F_3^{np}\left(l\right)
- 2 F_2^p\left(l\right) + 2 F_2^{np}\left(l\right)\right] .
\end{eqnarray}
We listed the contributions which involve $\phi_i$ and $a_{\alpha}$
exchanges separately.

The fermionic contribution in the 3d model is
\beq
{ 1\over f^4} \left[ -3F_3^p\left(l\right) +3F_3^{np}\left(l\right)
- F_2^p\left(l\right) +F_2^{np}\left(l\right) \right] .
\eeq

\section*{2-loop effective action}

By combining these contributions, we find the total 2-loop free energy
$F(l)$ of $U(1)$ NC gauge theory on fuzzy sphere as follows
\beqa
-F(l)&=&-36{1\over f^4}(F_1^p(l)-F_1^{np}(l)
+2F_2^p(l)-2F_2^{np}(l)
-F_3^p(l)+F_3^{np}(l))\n
&=&36{1\over f^4}(F_3^p(l)-F_3^{np}(l)) .
\eeqa
It is because we can prove that $F_1^p(l)+2F_2^p(l)=
F_1^{np}(l)+2F_2^{np}(l)=F_5(l)$ by using the Jacobi-identity.
In the case of 3d model, our result is
\beqa
-F_{3d}(l)&=&-{1\over f^4}(F_1^p(l)-F_1^{np}(l)
+2F_2^p(l)-2F_2^{np}(l)
-F_3^p(l)+F_3^{np}(l))\n
&=&{1\over f^4}(F_3^p(l)-F_3^{np}(l) ).
\eeqa

For $U(n)$ gauge group, we obtain
\beqa
-F(l)&=&-36{1\over f^4}\left(n^3 (F_1^p(l)+2F_2^p(l)
-F_3^p(l)+{1\over 4}F_5(l))
-n (F_1^{np}(l)+2F_2^{np}(l)-F_3^{np}(l)+{1\over 4}F_5(l))\right) \n
&=&-36{1\over f^4}\left(n^3 (
-F_3^p(l)+{5\over 4}F_5(l))
-n (-F_3^{np}(l)+{5\over 4}F_5(l))\right) ,
\eeqa
where
\beq
F_5(l)={1\over 2l+1}\sum_{j=1}^{2l}{2j+1\over (j(j+1))^2}
\rightarrow {1\over 2l+1} ~~(l\rightarrow \infty) .
\label{F5}
\eeq
$F_5(l)$ takes care of the absence of the $SU(2)$ singlet
state in the gluon propagator.
In the 3d model, the two loop effective action for $U(n)$ gauge group is
\beqa
-F_{3d}(l)&=&-{1\over f^4}\left(n^3 (F_1^p(l)+2F_2^p(l)-F_3^p(l)
+2F_5(l))
-n (F_1^{np}(l)+2F_2^{np}(l)-F_3^{np}(l)+2F_5(l))\right)\n
&=& -{1\over f^4}\left(n^3 (-F_3^p(l)
+3F_5(l))
-n (-F_3^{np}(l)+3F_5(l))\right).
\eeqa


\newpage

\begin{thebibliography}{99}
\bibitem{BFSS}
T. Banks, W. Fischler, S.H. Shenker and L. Susskind,
{\em M-theory as a matrix model: a conjecture},
Phys. Rev. {\bf D55} 5112 (1997), hep-th/9610043.
\bibitem{IKKT}N. Ishibashi, H. Kawai, Y. Kitazawa and A. Tsuchiya,
{\em A Large-N Reduced Model as Superstring},
Nucl. Phys. {\bf B498} (1997) 467, hep-th/9612115.
\bibitem{CDS} A. Connes, M. Douglas and A. Schwarz,
{\em Noncommutative geometry and matrix theory:
compactification on tori},
JHEP 9802: 003.1998, hep-th/9711162.
\bibitem{AIIKKT}
H. Aoki, N. Ishibashi, S. Iso, H. Kawai, Y. Kitazawa
and T. Tada,
{\em Non-commutative Yang-Mills in IIB matrix model},
Nucl. Phys. {\bf 565} (2000) 176,
hep-th/9908141.
\bibitem{Li}
M. Li, {\em Strings from IIB Matrices},
Nucl.Phys. {\bf B499} (1997) 149,hep-th/961222.
\bibitem{SW}  N. Seiberg and E. Witten,
{\em String theory and non-commutative geometry},
JHEP {\bf 9909} (1999) 032,
hep-th/9908142.
\bibitem{MRS}
S. Minwalla, M.V. Raamsdonk and N. Seiberg,
{\em Non-commutative Perturbative Dynamics},
JHEP {\bf 0002} (2000) 020,hep-th/9912072.
\bibitem{IIKK}
N. Ishibashi, S. Iso, H. Kawai and Y. Kitazawa, {\em Wilson loops in
non-commutative Yang-Mills}, Nucl. Phys. {\bf B573} (2000) 573,
hep-th/9910004.
\bibitem{Gross}
D. J. Gross, A. Hashimoto and N. Itzhaki,
{\em Observables of Non-Commutative Gauge Theories},
Adv.Theor.Math.Phys. {\bf 4} (2000) 893,hep-th/0008075.
\bibitem{DK}
A. Dhar and Y. Kitazawa,
{\em Non-commutative Gauge Theory, Open Wilson Lines and Closed Strings},
JHEP {\bf 0108} (2001) 044, hep-th/0106217.
\bibitem{MN}
K. Murakami and T. Nakatsu,
{\em Open Wilson Lines as States of Closed String},
hep-th/0211232.
\bibitem{Myers}
R.C. Myers,{\em Dielectric-Branes},
JHEP 9912 (1999) 022,hep-th/9910053.
\bibitem{Alekseev}
A. Y. Alekseev, A. Recknagel and V. Schomerus,
{\em Brane Dynamics in Background Fluxes and noncommutative Geometry},
JHEP {\bf 0005} (2000) 010, hep-th/0003187.
\bibitem{Mathom}
Y. Kitazawa,
{\em Matrix Models in Homogeneous Spaces},
Nucl Phys. {\bf B642} (2002) 210,
hep-th/0207115.
\bibitem{Masuda}
S. Aoyama and T. Masuda,
{\em The Fuzzy K\"{a}hler Coset Space with the Darboux Coordinates},
Phys. Lett. {\bf B 521} (2001) 376,hep-th/0109020.
\bibitem{IKK}
S. Iso, H. Kawai and Y. Kitazawa,
{\em Bi-local Fields in Non-commutative Field Theory},
Nucl.Phys. {\bf B576} (2000) 375,hep-th/0001027.
\bibitem{IKTW}
S. Iso, Y. Kimura, K. Tanaka and K. Wakatsuki,
{\em Non-commutative Gauge Theory on Fuzzy Sphere
from Matrix Model},
Nucl Phys. {\bf B604} (2001) 121,
hep-th/0101102.
\bibitem{Bonel}
G. Bonelli, {\em Matrix Strings in pp-wave backgrounds from
deformed Super Yang-Mills Theory},
JHEP 0208 (2002) 022; hep-th/0205213.
\bibitem{Kitazawa}
Y. Kitazawa,
{\em Vertex Operators in IIB Matrix Model},
JHEP {\bf 0204} (2002) 004,
hep-th/0201218.
\bibitem{deSit}
E. Witten,
{\em Quantum gravity in de Sitter space},
in Proceedings of Strings 2001,  eds. A. Dabholker, S. Mukhi and S. Wadia,
Clay Mathematics Institute (2002).
\bibitem{KNS}
W. Krauth, H. Nicolai and M. Staudacher,
{\em Monte Carlo Approach to M-Theory},
Phys. Lett. {\bf B431}, 31 (1998);
hep-th/9803117.
\bibitem{MNS}
G. W. Moore, N. Nekrasov and S. Shatashvili,
{\em D-particle bound states and generalized instantons},
Commun. Math. Phys. {\bf 209}, 77 (2000);
hep-th/9803265.
\bibitem{HLW}
C. Holzhey, F. Larsen and F. Wilczek,
{\em Geometric and Renormalized Entropy in Conformal Field Theory},
Nucl. Phys. {\bf B424}, 443 (1994); hep-th/9403108.
\bibitem{AK}
T. Aida and Y. Kitazawa,
{\em Quantum Gravity with Boundaries near Two Dimensions},
Mod. Phys. Lett. {\bf A10} 1351 (1995) ;hep-th/9504075.
\bibitem{BMN}
D. Berenstein, J. Maldacena and H. Nastase,
{\em Strings in flat space and pp waves from \cal{N}=4
super Yang Mills}, JHEP {\bf 0204} (2002) 013,
hep-th/0202021.
\bibitem{DSR}
K. Dasgupta, M. M. Sheikh-Jabbari, M. Van Raamsdonk,
{\em Matrix Perturbation Theory For M-Theory On a PP-wave},
JHEP {\bf 0205} (2002) 056,
hep-th/0205185.
\bibitem{Edm}
A. R. Edmonds,
{\em Angular Momentum in Quantum Mechanics},
Princeton Univ. Press (1957).
\bibitem{NS}
J. Nishimura and F. Sugino,
{\em Dynamical Generation of Four-Dimensional Space-Time
in IIB Matrix Model},
JHEP {\bf 0205} (2002) 001,
hep-th/0111102.
\bibitem{Kyoto}
H. Kawai, S. Kawamoto, T. Kuroki, T. Matsuo, S. Shinohara,
{\em  Mean Field Approximation of IIB Matrix Model and Emergence of Four
Dimensional Space-Time},Nucl.Phys. B647 (2002) 153-189:
hep-th/0204240;
{\em Improved Perturbation Theory and Four-Dimensional Space-Time in IIB
Matrix Model},Prog.Theor.Phys. 109 (2003) 115-132:hep-th/0211272 .
\end{thebibliography}
\end{document}